\documentclass[prd,amssymb,twocolumn,nofootinbib,10pt]{revtex4-1}
\usepackage{amsmath,amsfonts,amssymb}
\usepackage{color,verbatim,graphicx,float}

\begin{document}

\newtheorem{lemat}{Lemma}

\title{Nonlinear Field Space Cosmology} 

\author{Jakub Mielczarek$^{*}$}
\author{Tomasz Trze\'{s}niewski${}^{\dagger,*}$}

\affiliation{${}^*$Institute of Physics, Jagiellonian University, ul.\ \L{}ojasiewicza 11, 
30-348 Krak\'{o}w, Poland \\
${}^\dagger$Institute for Theoretical Physics, University of Wroc\l{}aw, pl.\ 
Borna 9, 50-204 Wroc\l{}aw, Poland
}

\begin{abstract}
We consider the FRW cosmological model in which the matter
content of universe (playing a role of inflaton or quintessence) is given by a novel generalization 
of the massive scalar field. The latter is a scalar version of the recently introduced Nonlinear 
Field Space Theory (NFST), where physical phase space of a given field is assumed to be 
compactified at large energies. For our analysis we choose the simple case of a field with the 
spherical phase space and endow it with the generalized Hamiltonian analogous to the XXZ 
Heisenberg model, normally describing a system of spins in condensed matter physics. 
Subsequently, we study both the homogenous cosmological sector and linear perturbations 
of such a test field. In the homogenous sector we find that nonlinearity of the field phase 
space is becoming relevant for large volumes of universe and then it can lead to a recollapse,
and possibly also at very high energies, leading to the phase of a bounce.
Quantization of the field is performed in the limit where nontrivial nature of its phase space 
can be neglected, while there is a non-vanishing contribution from the Lorentz symmetry 
breaking term of the Hamiltonian. As a result, in the leading order of the XXZ anisotropy 
parameter, we find that the inflationary spectral index remains unmodified with respect to the 
standard case but the total amplitude of perturbations is subject to a correction.  
The Bunch-Davies vacuum state also becomes appropriately corrected. The proposed 
new approach is bringing cosmology and condensed matter physics closer together, which 
may turn out to be beneficial for both disciplines.
\end{abstract}  


\maketitle

\section{Introduction}

Scalar fields play a significant role in our understanding of the cosmological dynamics and structure 
formation. Both the inflationary epoch and the current phase of dark energy domination can be 
modeled using the scalar matter (an inflaton or quintessence field). On the other hand, the increasingly 
more precise astronomical observations provide the opportunity to test various modifications of the ordinary 
scalar field theories. The modified models turn out to be useful because the standard approach is insufficient to 
explain all subtleties in the observational data. In particular, the explored research directions include String 
Theory-inspired cosmological models, such as the Dirac-Born-Infeld \cite{Silverstein:2003hf,Alishahiha:2004eh} 
and multifield models \cite{Gong:2011uw}.  Furthermore, recent results obtained within loop quantum 
cosmology \cite{Bojowald:2008zzb,Ashtekar:2011ni} provide novel solutions to the problems of classical 
cosmology, which can also lead to rethinking of the established methods. 

The Dirac-Born-Infeld and multifield models share certain features with the 
\emph{Nonlinear Field Space Theory} (NFST) \cite{Mielczarek:2016rax,Trzesniewski:2017lpb},
which has recently been proposed in a broad theoretical context. The essential idea of the NFST 
framework is rather simple. Namely, standard linear (i.e.\! affine) phase space of a given field theory 
is considered to be only a local approximation of some more general (nonlinear) field phase space, 
which also leads to a generalization of the field Hamiltonian. The freedom in the choice of nontrivial 
topology or geometry of phase spaces is in principle very large. For the mathematical and physical 
consistency it is enough to require that the nonlinear phase space remains a symplectic manifold, 
while the ordinary field theory is completely recovered in the appropriate limit. We also have to 
mention that field theories known as the non-linear sigma models, especially in the Tseytlin formulation 
\cite{Freidel:2014qy}, can be seen as a special type of NFST, although in the approach of 
\cite{Freidel:2014qy} they are defined on string worldsheets rather than spacetime and their 
target spaces are not Riemannian. 

One of the advantages of NFST is that it allows to naturally implement the ``Principle of finiteness'',
which was the original motivation behind the Born-Infeld theory \cite{Born:1934}. In the Born-Infeld 
theory the time derivative of a field is constrained analogously to the velocity of a relativistic particle. 
Besides, from the modern, string-theoretic point of view, the value of a Born-Infeld scalar field is 
associated with the distance between branes in the higher dimensional space \cite{Silverstein:2003hf,
Alishahiha:2004eh}. The required conservation of the causal structure leads to a constraint on the 
velocity with which a brane can move, which translates into the corresponding constraint on the velocity 
(i.e.\! rate of change) of the Born-Infeld field. The restriction implies that the field Lagrangian is given 
by the Lorentz factor for the field velocity. On the other hand, in the NFST framework the allowed field 
values (as well as their conjugate momenta) are determined by the nontrivial structure of an assumed 
nonlinear phase space. On the latter we have to find the Hamiltonian that appropriately generalizes a 
given standard field theory. 

The defining feature of NFST is the nonlinearity of field phase spaces. In this sense multi-field models, 
where the configuration subspace of field phase space usually has non-vanishing curvature, can be 
seen as a special subclass of NFST. However, a NFST model can also be constructed from a 
single-component scalar field, as we will discuss in a moment. In the case of multi-field models the nontrivial 
geometry is not introduced in the field's momentum space, which ensures that the Lorentz symmetry of the field 
is preserved. In turn, in NFST the whole phase space is generally 
non-trivial and this may not even allow a global decomposition of phase space into the configuration 
and momentum spaces. While symmetries of the background spacetime are not modified, the Lorentz 
symmetry of the field will usually be violated.  

One still might ask whether there are any strong reasons to generalize the standard 
linear phase spaces of fields. A number of arguments and motivations, related especially 
to quantum gravity, can be found in the original reference \cite{Mielczarek:2016rax}. 
Let us here mention two important premises. The first one is related to special properties 
of compact phase spaces. Namely, when the area of phase space is finite, it can 
accommodate only a finite number of degrees of freedom, each occupying a cell with 
the area of $2\pi \hbar$.\footnote{The area may change if the uncertainty relation is deformed.} 
This implies the finite dimensionality of the Hilbert space in the corresponding quantum 
theory (see below), which is the quantum counterpart of the Principle of finiteness and may 
naturally solve the problem of UV divergences in quantum field theory. The second premise 
concerns the particular example of a compact phase space that is given by a sphere and 
therefore is equivalent to the phase space of a spin. The latter relation leads to an exact analogy 
between fields (with compact phase spaces) and spin systems, first observed in \cite{Mielczarek:2016xql}, 
where it has been called the Spin-Field Correspondence. In principle, this kind of duality should allow us to 
find the spin system counterparts for different types of field theories. 
Such an approach is in the spirit of the broadly considered analog condensed 
matter models of gravity \cite{Barcelo:2005fc}. However, it should not 
only allow to design a condensed matter system that, in the linear limit, emulates a given 
field theory, but also to hypothesize about the fundamental origin of physical fields.  

In order to introduce NFST in the cosmological context we first have to discuss a scalar 
field theory defined on Minkowski spacetime. In such a case phase space of the field at 
any point of space, or for any Fourier mode, is a linear space $\mathbb{R}^2$. 
Then the simplest construction of a compact NFST model is to reinterpret the field phase space 
$\mathbb{R}^2$ as the small field approximation of a sphere $S^2$ that is equipped with the 
appropriate symplectic form and Hamiltonian. In \cite{Mielczarek:2016rax} 
(see also \cite{Trzesniewski:2017lpb}) it has been done at the level of Fourier 
modes and in \cite{Mielczarek:2016xql} at the level of space points, which gives us two 
inequivalent models where the total field phase space is an infinite collection of spheres. 
In this paper we will restrict to the model constructed in the position representation, since 
it has a direct connection with spin systems in condensed matter physics. 

\begin{figure}[ht!]
\centering
\includegraphics[width=10cm,angle=0]{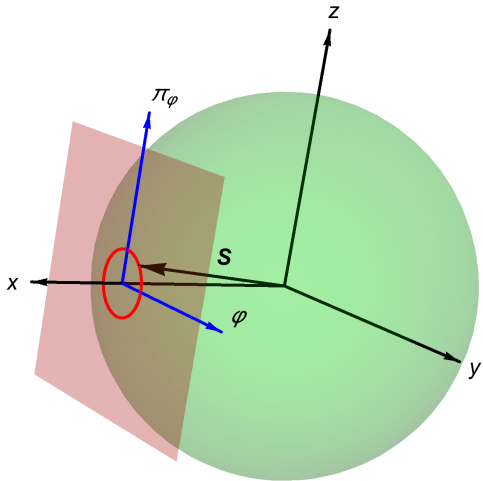}
\caption{Geometrical relation between a vector ${\bf S}$ and variables $\varphi,\pi_\varphi$ in the 
linear approximation. The precession of ${\bf S}$ around the $x$ axis (see below) determines a 
closed trajectory in the $(\varphi,\pi_\varphi)$ linear phase space, which is an ellipse for small 
precession angles.} 
\label{SF}
\end{figure}

There are three parametrizations of the spherical phase space that will be of interest to us here (see Fig.~\ref{SF}). 
The first one is given by usual angular coordinates $\phi \in [-\pi,\pi)$ 
and $\theta \in [0,\pi]$. The second uses the physical phases space variables $\varphi$ and 
$\pi_\varphi$, with the origin at $(\phi,\theta) = (0,\pi/2)$. More details on the description of phase space in this 
parametrization can be found in \cite{Mielczarek:2016rax,Mielczarek:2016xql}. 
Finally, the third possibility is the Cartesian parametrization, in which coordinates represent components 
of a spin vector ${\bf S} = (S_x,S_y,S_z)$ and in terms of the 
two other parameterizations they can be expressed as:
\begin{align}
S_x &:= S \sin\theta \cos\phi = S \cos\left(\frac{\pi_{\varphi}}{R_2}\right) \cos\left(\frac{\varphi}{R_1}\right), 
\label{Sx} \\
S_y &:= S \sin\theta \sin\phi = S \cos\left(\frac{\pi_{\varphi}}{R_2}\right) \sin\left(\frac{\varphi}{R_1}\right), 
\label{Sy}\\
S_z &:= S \cos\theta = S \sin\left(\frac{\pi_{\varphi}}{R_2}\right), 
\label{Sz}
\end{align}
together with the obvious condition $S_x^2 + S_y^2 + S_z^2 = S^2$, where $S$ is the sphere's radius. 
Moreover, the dimensionful constants $R_1$ and $R_2$ satisfy the relation $R_1 R_2 = S$, 
so that the standard symplectic form on a sphere in terms of the $\varphi$ and $\pi_\varphi$ 
variables is given by
\begin{align}
\omega = \cos\left(\frac{\pi_\varphi}{R_2}\right) d\pi_{\varphi} \wedge d\varphi\,. 
\label{symplectic1}
\end{align}
Since $\omega$ is equal to the area two-form, integrating it over the whole phase space one obtains 
the total area of the latter: ${\rm Ar}(S^2) = \int_{S^2} \omega = 4\pi S$. Meanwhile, in the ordinary 
quantum theory every degree of freedom occupies the area $2\pi \hbar$ and hence we infer that $S$ 
is subject to the quantization condition $S = j\hslash$, with $2j \in \mathbb{N}$. This allows us to relate 
dimension of the spin Hilbert space $\text{dim}({\cal H}_j) = 2j + 1$ with the area of 
phase space in the semiclassical regime. 

We now need to choose the Hamiltonian on the discussed phase space. According to what we mentioned 
above, a spherical phase space of the field, which is attached at each point of space, can also be seen as 
describing a fictitious spin. Therefore, we may try to apply here the known models from condensed matter 
physics. In this paper we will restrict our investigations to the XXZ generalization of the (continuous) Heisenberg 
model. As discussed in \cite{Mielczarek:2016xql}, the XXX Heisenberg model coupled to a constant magnetic 
field emulates the non-relativistic scalar NFST with the quadratic dispersion relation. On the other hand, it has 
been shown \cite{XYZSF} that the generalization of the Heisenberg XXX model to the XXZ model allows us to 
recover the relativistic scalar NFST, in the limit of vanishing anisotropy parameter $\Delta$. 

Let us first write the Hamiltonian for the discrete XXZ 
Heisenberg model of spins on a cubic lattice in three spatial dimensions:
\begin{align}
H_{XXZ} &= -J \sum_{i,j} \left(S_x^{(i)} S_x^{(j)} + S_y^{(i)} S_y^{(j)} + \Delta S_z^{(i)} S_z^{(j)}\right) \nonumber \\
&- \mu \sum_i {\bf B} \cdot {\bf S }^{(i)}\,, 
\label{XXZDis}
\end{align}
where the first sum is performed over the nearest neighbors. $J$ and $\mu$ denote 
the coupling constants, ${\bf B}$ is an external magnetic field and $\Delta$ the
dimensionless anisotropy parameter, defined so that for $\Delta = 1$ the XXZ Heisenberg 
model reduces to the XXX model. The interaction of spins ${\bf S}^{(i)}$ with the 
magnetic field ${\bf B}$ leads to the spin precession, which plays a crucial role 
in the duality between a spin system and NFST that has been mentioned above (see also the next Section). 
In the continuum limit the Hamiltonian (\ref{XXZDis}) becomes
\begin{align}
H_{XXZ}^{\text{cont}} &= -\tilde{J} \int d^3x \left[(\nabla S_x)^2 + (\nabla S_y)^2 + \Delta (\nabla S_z)^2\right] \nonumber \\
&- \tilde{\mu} \int d^3x\, {\bf B} \cdot {\bf S}\,, 
\label{XXZCont}
\end{align}
with the new coupling constants $\tilde{J}$ and $\tilde{\mu}$. It should be stressed that in the continuous 
case the vector ${\bf S}$ naturally gains the dimension of density (i.e.\! $1/[{\rm length}]^3$ in 
the units of $\hbar = 1 = c$). However, for the later convenience the dimension can be absorbed into 
the definitions of $\tilde{J}$ and $\tilde{\mu}$. 

There are two main objectives of this paper. Firstly, to study an application of the scalar NFST as an inflaton 
or quintessence field on the FRW background. Secondly, to examine a possibility of testing predictions of the 
considered framework with the use of present astronomical observations. In Sec.~\ref{Hom} we start by 
defining a homogenous cosmological model with a matter field corresponding to the continuous XXZ 
Heisenberg model. Its dynamics is analyzed in the Hamiltonian framework for an arbitrary value of $S$. 
Subsequently, in Sec.~\ref{Pert} we derive the leading order inhomogeneities, which correspond to the 
limit $S \rightarrow \infty$. The Hamiltonian obtained from the XXZ model includes the Lorentz symmetry breaking term, 
proportional to the anisotropy parameter $\Delta$. We study generation of quantum inflationary inhomogeneities 
taking this effect into account. Finally, in Sec.~\ref{Sum} we summarize our results and discuss prospects 
of the further development of the considered framework.

\section{Homogeneous cosmological model} \label{Hom}

The purpose of this Section is to construct a homogeneous and isotropic cosmological model with 
the scalar matter field described by the XXZ Heisenberg model (\ref{XXZCont}). According to the 
results of \cite{Mielczarek:2016xql}, one can expect that the Hamiltonian (\ref{XXZCont}) in the 
leading order is equivalent to the Hamiltonian of a massive scalar field on the FRW background, 
which is widely used to model the inflationary dynamics. Below we will relate the parameters appearing 
in the Heisenberg model with the scalar field's mass and the cosmological scale factor, so that in the 
limit $S \rightarrow \infty$ we can recover the known expressions for an ordinary inflationary model. 
To this end let us begin with the case of a standard scalar field defined on the FRW background. 

\subsection{Symplectic form}

Analogously to the case of Minkowski spacetime discussed in the Introduction, we first note that 
phase space of a scalar field on the FRW background is $\mathbb{R} \times \mathbb{R}$ equipped 
with the symplectic form $\omega_{\varphi} = V_0 d\pi_\varphi \wedge d\varphi$ (here $\varphi,\pi_\varphi$ 
are coordinates on $\mathbb{R}^2$). The factor $V_0$ denotes a fiducial volume over which the 
Hamiltonian density is integrated and we introduce it since the integration over the whole spatial 
slice $\mathbb{R}^3$ leads to an infinite result (i.e.\! IR divergence). Thereafter, $V_0$ will be 
absorbed into the definition of field variables and will not appear in the final results. Namely, the 
momentum $\pi_\varphi$ can be rescaled as follows $\pi_\varphi \rightarrow \pi_\varphi/V_0$, so that 
the symplectic form for the scalar field becomes $\omega_\varphi = d\pi_\varphi \wedge d\varphi$.
 
The other ingredient of an ordinary inflationary model is the dynamical FRW background. Its degrees 
of freedom are usually chosen to be the dimensionless scale factor $a$, which relates the comoving 
and physical distances, and the conjugate momentum. The scale factor can be normalized in such a way 
that $a = 1$ at some chosen moment of time. For convenience, in this paper we replace $a$ by the 
dimensionful volume variable $q \equiv V_0 a^3$, which measures the expanding (or contracting) 
volume of a region $V_0$. $q$ is complemented by the canonically conjugate momentum $p$. 

Consequently, the total phase space of the model is four dimensional and its symplectic form 
is assumed to be composed of two independent contributions, namely
\begin{align}
\omega_{\text{tot}} = \omega_{\text{FRW}} + \omega_\varphi\,, \label{SympStandard}
\end{align}
where $\omega_\varphi = d\pi_\varphi \wedge d\varphi$ and the gravitational component has the canonical form
\begin{align}
\omega_{\text{FRW}} = dp \wedge dq\,.
\end{align}

Now we would like to generalize the standard symplectic form (\ref{SympStandard}) to the case of 
a scalar field with the spherical phase space. 
The symplectic form for such a field defined on the Minkowski background is given by (\ref{symplectic1}). 
Therefore, our straightforward approach for the FRW background is to consider the following two-form
\begin{align}
\omega = dp \wedge dq + \cos\left(\frac{\pi_\varphi}{R_2(q)}\right) d\pi_\varphi \wedge d\varphi\,, 
\label{SympNew}
\end{align}
where we allow the $R_2$ constant to become some function of $q$. In the 
$R_2 \rightarrow \infty$ limit the form (\ref{SympNew}) reduces to the symplectic form 
(\ref{SympStandard}), as it should be for the model in the linear phase space limit. The basic requirement 
of the correspondence with the standard scalar field is, therefore, satisfied. 

The second requirement is that the two-form (\ref{SympNew}) is symplectic, which is satisfied if and 
only if it is closed (i.e.\! $d\omega = 0$). In the case of the canonical form (\ref{SympStandard}) this 
condition is trivial. On the other hand, for the new symplectic form (\ref{SympNew}) it implies 
that
\begin{align}
R_2(q) = R_2 = {\rm const}\,, 
\end{align}  
as in (\ref{symplectic1}). The matter contribution to the total symplectic form is therefore 
expected to be independent of the variable $q$. Actually, this is analogous to an ordinary 
scalar field theory, in which passing from the Minkowski to FRW background does not affect the form 
$\omega_\varphi$. 

Inverting the symplectic form (\ref{SympNew}) we can derive the corresponding Poisson bracket
\begin{align}
\left\{\cdot,\cdot\right\} &:= (\omega^{-1})^{ij}(\partial_i \cdot )(\partial_j \cdot ) = \left[ \frac{\partial\cdot}{\partial q} \frac{\partial\cdot}{\partial p} 
- \frac{\partial\cdot}{\partial p} \frac{\partial\cdot}{\partial q} \right]  \nonumber \\
&+ \frac{1}{\cos\left(\frac{\pi_\varphi}{R_2}\right)} \left[ \frac{\partial\cdot}{\partial\varphi} 
\frac{\partial\cdot}{\partial\pi_\varphi} 
- \frac{\partial\cdot}{\partial\pi_\varphi} \frac{\partial\cdot}{\partial\varphi} \right], 
\label{Poisson1}
\end{align}
which will be used in the further analysis of the model's dynamics. Nevertheless, one might still wonder whether a different 
generalization of the standard symplectic form (\ref{SympStandard}) to the case with the spherical phase space, leading to 
a different bracket than (\ref{Poisson1}), could also give us valuable results. In the Appendix we briefly discuss one such possibility, 
which is obtained by relaxing the requirement that the two-form (\ref{SympNew}) is closed. The two-form is still constrained by 
the topology of phase space and has to be closely related to (\ref{symplectic1}). In particular, a function other than cosine would 
either naturally correspond to a different topology or not lead to a sensibly defined phase space.

\subsection{Matter Hamiltonian}

Dynamics of an ordinary homogeneous massive scalar field is determined by the Hamiltonian
\begin{align}
H_\varphi &= \int_{V_0} d^3x\, N a^3 \left(\frac{\pi^2_\varphi}{2a^6} + \frac{1}{2} m^2 \varphi^2\right) \nonumber \\
&= N V_0 a^3 \left(\frac{\pi^2_\varphi}{2a^6} + \frac{1}{2} m^2 \varphi^2\right), 
\label{HamStandard}
\end{align}
where $m$ is the field's mass and $N$ denotes the lapse function, which is associated with the time 
reparametrization symmetry. Due to the homogeneity of the integrand in (\ref{HamStandard}), the 
integration reduces to $\int_{V_0} d^3x = V_0$. With the use of the $q$ variable introduced in the 
previous Subsection and the rescaling $\pi_\varphi \rightarrow \pi_\varphi/V_0$ (necessary to obtain 
the well normalized symplectic form $\omega_\varphi = d\pi_\varphi \wedge d\varphi$), one can simplify the 
Hamiltonian (\ref{HamStandard}) to
\begin{align}
H_{\varphi} = N q \left(\frac{\pi_\varphi^2}{2q^2} + \frac{1}{2} m^2 \varphi^2\right). 
\label{HamStandard2}
\end{align}
This is the expression that should be recovered in the $S \rightarrow \infty$ limit of a homogenous scalar 
NFST, which is described by the Hamiltonian analogous to the XXZ Heisenberg model (\ref{XXZCont}). 

The first term of (\ref{XXZCont}) contains spatial derivatives and therefore will not contribute 
to the homogenous sector of the corresponding scalar field. What matters is the term describing 
interaction with a magnetic field ${\bf B}$. Similarly as in \cite{Mielczarek:2016rax,Mielczarek:2016xql}, 
we choose the vector ${\bf B}$ to be oriented along the $x$ axis (i.e.\! ${\bf B} := (B_x,0,0)$), 
so that the precession of a spin ${\bf S}$ occurs around this direction (as depicted in Fig.~\ref{SF}). Then 
to obtain the scalar field Hamiltonian corresponding to $-\tilde{\mu} \int d^3x\, B_x S_x$ we have 
to introduce the cosmological scaling and freedom of time reparametrization, which can be 
accomplished through multiplying the measure $d^3x$ by $N a^3$. The resulting Hamiltonian is
\begin{align}
H_{{\bf S}} &= -N q \tilde{\mu} B_x S_x \label{HamSpin1}\\
&= N q \left(-\tilde{\mu} B_x S + \frac{\tilde{\mu} B_x S}{2R_2^2} \pi^2_\varphi 
+ \frac{\tilde{\mu} B_x S}{2R_1^2} \varphi^2 + {\cal O}(4)\right), \nonumber
\end{align}
where we subsequently expanded the spin component $S_x$ (given by the expression (\ref{Sx})) 
up to the quadratic order in the field variables $\varphi$, $\pi_\varphi$. 

While we have chosen $R_2 = {\rm const}$, $R_1$ can still depend on $q$. Moreover, 
for $q = V_0 \equiv q_0$ (i.e.\! $a = 1$), which corresponds to the Minkowski background, 
the condition $R_1 R_2 = S$ has to be satisfied \cite{Mielczarek:2016xql}. Taking these 
issues into account, the comparison of (\ref{HamSpin1}) with (\ref{HamStandard2}) allows 
us to identify the following relations between the parameters of both Hamiltonians:
\begin{align}
\tilde{\mu} B_x &= \left(\frac{q_0}{q}\right) \frac{m}{q}\,, \label{massed}\\ 
R_1 &= \frac{1}{q} \sqrt{\frac{S q_0}{m}}\,, \label{R1def}\\ 
R_2 &= \sqrt{S q_0 m}\,, \label{R2def}
\end{align}
and consequently $R_1 R_2 = \frac{q_0}{q} S$. The value of $q_0$ could in principle 
be set to $q_0 = 1$ but we will leave it unspecified in order to keep track of the dimensions. 

Finally, we note that the extra term $-\tilde{\mu} B_x S = -S m q_0/q^2$ in the Hamiltonian 
(\ref{HamSpin1}) can be eliminated by subtracting the constant $S$ from $S_x$, which leads 
to the properly normalized scalar field Hamiltonian of the form
\begin{align}
H_{\bf S} &= N m \left(\frac{q_0}{q} \right) \left(S - S_x\right) \nonumber\\ 
&= N q \left(\frac{\pi^2_\varphi}{2q^2} + \frac{1}{2} m^2 \varphi^2\right) + {\cal O}(4)\,. 
\label{HamSpin2}
\end{align}
The $q$-dependent energy shift in (\ref{HamSpin2}) ensures that energy is positive definite 
and the Hamiltonian $H_{\bf S}$ vanishes at the classical minimum $(\varphi,\pi_\varphi) = (0,0)$. 
This is analogous to the theory on the Minkowski background \cite{Mielczarek:2016rax}. 

Therefore, an ordinary massive scalar field is indeed recovered for small field values from the NFST 
Hamiltonian (\ref{HamSpin1}) analogous to the XXZ Heisenberg model. 
Let us again stress that (\ref{HamSpin1}) can be introduced as a generalization 
of the scalar field Hamiltonian (\ref{HamStandard2}) due to the assumed spherical 
geometry of the field's phase space at every point of spacetime, which is mathematically identical to the 
phase space of a spin. Furthermore, the form of the scalar field's potential is determined by an interaction 
of such (probably) fictitious spins with a constant vector field that, in the context of condensed matter, 
plays a role of the external magnetic field. 

On the other hand, if the negative term in the expansion of (\ref{HamSpin1}) is not subtracted, the 
Hamiltonian (\ref{HamSpin2}) becomes replaced by 
\begin{align}
\tilde{H}_{\bf S} = H_{\bf S} + \delta H_{\bf S} = -N m \left(\frac{q_0}{q}\right) S_x\,,
\end{align}
which is negative in the regime $S_x > 0$, where the standard 
limit should be recovered. Furthermore, the $\delta H_{\bf S}$ term is always negative and 
can be perceived as a source of negative energy, with the energy density
\begin{align}
\rho_* = -S m \frac{q_0}{q^2}\,. 
\label{rhonneg}
\end{align}   
This contribution scales as $1/a^6$ and therefore can play a dominant role in the early universe, 
while becoming irrelevant for the late time dynamics. In particular, it may lead to the 
phase of non-singular bounce, which replaces the big bang singularity 
(see Subsection \ref{Dynamics}).

\subsection{Equations of motion}

According to the previous Subsections, the total Hamiltonian for the system under consideration is
\begin{align}
H_{\text{tot}} &= H_{\text{FRW}} + H_{{\bf S}} \nonumber\\ 
&= N q \left(-\frac{3}{4} \kappa p^2 + \frac{m}{q} \left(\frac{q_0}{q} \right)  \left(S - S_x\right) \right), 
\label{TotalHam}
\end{align}
where $\kappa \equiv 8\pi G = 8\pi/m_{\text{Pl}}^2$, $m_{\text{Pl}}$ is the Planck mass 
and the matter field Hamiltonian is given by (\ref{HamSpin2}). Using (\ref{TotalHam}) 
one can derive the Hamilton equations $\dot{f} = \{f, H_{\text{tot}}\}$ for an arbitrary function 
$f$ on phase space $(q,p,\varphi,\pi_\varphi)$. On the other hand, the Hamiltonian 
(\ref{TotalHam}) can also be seen as a constraint, imposed through the condition 
$\frac{\partial}{\partial N} H_{\text{tot}} = 0$ (which is equivalent to vanishing of the conjugate 
momentum of $N$, i.e.\! $p_N = 0$). The constraint 
can be written as:
\begin{align}
\frac{3}{4} \kappa p^2 = \frac{m}{q} \left(\frac{q_0}{q} \right)  \left(S - S_x\right). 
\label{HamConstr}
\end{align} 
In the usual way we also introduce the Hubble factor
\begin{align}
H \equiv \frac{1}{3} \frac{\dot{q}}{q} = \frac{1}{3q} \left\{q, H_{\text{tot}} \right\} 
= -\frac{1}{2} \kappa p\,, \label{HubbleDef}
\end{align}
where we have chosen the 
gauge $N = 1$, which will be kept in the remaining part of this Section. In such a case the overdot ``$\ \dot{}\ $" denotes a 
differentiation with respect to the coordinate time $t$. 

Expressing $p$ via (\ref{HubbleDef}) and substituting it into the Hamiltonian constraint (\ref{HamConstr}) we obtain 
the Friedmann equation
\begin{align}
H^2 = \frac{1}{9} \left(\frac{\dot{q}}{q}\right)^2 = \frac{\kappa}{3} \rho\,, 
\label{FriedmannEq}
\end{align}
with the matter energy density
\begin{align}
\rho = \frac{m}{q} \left(\frac{q_0}{q} \right)  \left(S - S_x\right). 
\label{EnergyDensity}
\end{align} 
The density $\rho$ is positive definite and vanishes in the limit $S_x \rightarrow S$. 

Let us now derive the remaining equations of motion. We will do it for both the 
components of a spin vector ${\bf S} = (S_x,S_y,S_z)$ (as functions of $\varphi$ and $\pi_\varphi$) and the field variables 
$\varphi$ and $\pi_\varphi$. The advantage of using the $S_x,S_y,S_z$ variables 
is that, in contrast to $\varphi$ and $\pi_\varphi$, they are well defined on 
the whole $S^2$ phase space. As one can verify by a direct calculation, they naturally generate the $\mathfrak{so}(3)$ 
algebra
\begin{align}
\{S_x,S_y\} = S_z\,, \ \ \{S_z,S_x\} = S_y\,, \ \ \{S_y,S_z\} = S_x\,.
\end{align}
Moreover, we find that their Poisson brackets with the gravitational variables $q$ and $p$ have the form
\begin{align}
\{S_x,q\} = \{S_y,q\} = \{S_z,q\} = 0
\end{align}
and
\begin{align}
\{S_x,p\} &= \frac{\partial S_x}{\partial q} = -\frac{S_y}{q} \arctan\frac{S_y}{S_x}\,, \\ 
\{S_y,p\} &= \frac{\partial S_y}{\partial q} = \frac{S_x}{q} \arctan\frac{S_y}{S_x}\,, \\ 
\{S_z,p\} &= \frac{\partial S_z}{\partial q} = 0\,.
\end{align}
Using the above formulae we derive the evolution equations
\begin{align}
\dot{S}_x &= \{S_x, H_{\text{tot}}\} = \frac{3}{2} N \kappa p\, S_y \arctan\frac{S_y}{S_x}, \label{eqSx} \\ 
\dot{S}_y &= \{S_y, H_{\text{tot}}\} = N m\, S_z - \frac{3}{2} N \kappa p\, S_x \arctan\frac{S_y}{S_x}, \label{eqSy} \\ 
\dot{S}_z &= \{S_z, H_{\text{tot}}\} = -N m\, S_y\,, \label{eqSz}
\end{align}
and they naturally satisfy $\partial_t S^2 = 2(S_x \dot{S}_x + S_y \dot{S}_y + S_z \dot{S}_z) = 0$. We also find that the equation of motion for $p$ is
\begin{align}
\dot{p} &= \frac{3}{4} N \kappa p^2 \nonumber\\
&+ \frac{Nm}{q} \left(\frac{q_0}{q} \right) \left(S - S_x - S_y \arctan\frac{S_y}{S_x}\right). \label{eqp}
\end{align}
On the other hand, for the $\varphi$ and $\pi_\varphi$ variables we calculate
\begin{align}
\dot{\varphi} &= \frac{1}{\cos\left( \frac{\pi_\varphi}{R_2} \right)}
\frac{\partial H_{\text{tot}}}{\partial\pi_\varphi} \nonumber\\
&= N \frac{R_2}{q} \tan\left( \frac{\pi_\varphi}{R_2} \right) \cos\left( \frac{\varphi}{R_1} \right), \label{eqphi}\\
\dot{\pi}_\varphi &= -\frac{1}{\cos\left( \frac{\pi_\varphi}{R_2} \right)}
\frac{\partial H_{\text{tot}}}{\partial\varphi} = -N q R_1 m^2 \sin\left( \frac{\varphi}{R_1} \right), \label{eqpi}
\end{align}
which are well posed on the hemisphere where $\frac{\varphi}{R_1} \in (-\pi/2,\pi/2)$ and $\frac{\pi_\varphi}{R_2} \in (-\pi/2,\pi/2)$. 
In the case of Minkowski spacetime ($q = {\rm const}$), the exact solutions of the equations (\ref{eqphi}-\ref{eqpi}) can actually be found (see \cite{Trzesniewski:2017lpb}). 

In general, as one can easily verify, the standard equations of motion for a massive scalar field are recovered in the 
limit of $R_1 \rightarrow \infty$, $R_2 \rightarrow \infty$. Both of the limits are obtained when $S \rightarrow \infty$. 
However, one has to keep in mind that $R_1$ is actually a function of $q$. From (\ref{R1def}) we infer that in 
the large volume limit $q \rightarrow \infty$ we have $R_1 \rightarrow 0$. Therefore, while taking 
$S \rightarrow \infty$ always leads to the standard field dynamics, it is not obvious if this dynamics is 
recovered for some finite $S$ in the large volume limit, $q \rightarrow \infty$. This issue will be addressed in 
the next Subsection. 

\subsection{Basic features of the dynamics}  \label{Dynamics}

In this Subsection we present a preliminary discussion of the features of dynamics described by evolution 
equations calculated above. The complete analysis is beyond the scope of this paper and will be a subject 
of the future investigations. Here we focus on the most basic properties of the considered model. 

We start by deriving the ${\cal O}(1/S)$ corrections to the standard equations 
of motion expected for cosmology with a massive scalar field. To this end let us first remind that 
in the case of such an ordinary scalar field the expressions for its energy density and pressure respectively have the form
\begin{align} 
\rho_\varphi &:= \frac{\pi_\varphi^2}{2q^2} + \frac{1}{2} m^2 \varphi^2\,, \label{rhovarphi} \\ 
P_\varphi &:= \frac{\pi_\varphi^2}{2q^2} - \frac{1}{2} m^2 \varphi^2\,. \label{Pvarphi}
\end{align}
Applying the above definitions to the Friedmann equation (\ref{FriedmannEq}), where we expand the energy 
density (\ref{EnergyDensity}) as a series in $1/S$, we obtain
\begin{align}
H^2 = \frac{\kappa}{3} \rho_\varphi - \frac{\kappa}{9} \frac{q^2}{S q_0 m}
\left(\rho_\varphi^2 - \frac{1}{2} P_\varphi^2\right) + {\cal O}(1/S^2)\,. \label{ModFried}
\end{align}
As one can see, the leading order correction does not only depend on the field's energy 
density but also on its pressure. These new contributions become more and more relevant 
with increasing $q$. Therefore, we can expect that the spherical geometry of phase 
space modifies the late time dynamics. In particular, the correction in (\ref{ModFried}) 
becomes negative if $\rho_\varphi^2 - \frac{1}{2} P_\varphi^2 > 0$, and this can lead 
to the effect of recollapse, as we will discuss below. For the special case of the barotropic
equation of state $P_\varphi = w \rho_\varphi$, the correction term remains negative 
if the condition $|w|<\sqrt{2}$ is satisfied, which covers most of the types of matter 
considered in cosmology. 
  
Furthermore, let us observe that expanding the equations of motion (\ref{eqphi}-\ref{eqpi}) up to the 
first order in $1/S$ we obtain
\begin{align}
\dot{\varphi} &= \frac{\pi_\varphi}{q} + \frac{\pi_\varphi}{S m} \frac{q}{q_0} 
\left(\frac{\pi_\varphi^2}{3q^2} - \frac{m^2}{2} \varphi^2\right) + {\cal O}(1/S^2)\,, \label{varphieq1/S} \\ 
\dot{\pi}_\varphi &= -q m^2\, \varphi + \frac{q^3 m^3}{6S q_0} \varphi^3 + {\cal O}(1/S^2)\,. \label{pieq1/S}
\end{align}
Combining the above equations we can derive the modified Klein-Gordon equation
\begin{align}
\ddot{\varphi} + 3H \dot{\varphi} + m^2 \varphi &= 
-\frac{q^2 m}{S q_0} \left[ 3H \dot{\varphi} \varphi^2 + 2\dot{\varphi}^2 \varphi - \frac{2m^2}{3} \varphi^3 \right] \nonumber \\ 
&+ {\cal O}(1/S^2)\,, \label{KGeq}
\end{align}
which is quite complicated and therefore we do not analyze it further here. In the leading order, 
the late time oscillation of the field at the bottom of the potential well is approximated 
by the solutions:
\begin{align}
\varphi \propto \frac{\cos(m t + \alpha)}{t} \quad \text{and} \quad q \propto t^2\,,
\end{align} 
where $\alpha$ is a constant of integration. Hence one can show that at late times the average pressure  
is approximately zero $\langle P_\varphi \rangle \approx 0$, while energy density $\rho_\varphi \sim 1/q$. 
In other words, in this regime the field effectively behaves like a dust matter. 

Consequently, the modified Friedmann equation (\ref{ModFried}) at large $q$ (i.e.\! late times) 
can be written as
\begin{align}
H^2 \approx \frac{\kappa}{3} \rho_\varphi \left(1 - \frac{\rho_\varphi}{\rho_X}\right)\,, 
\label{H2effective}
\end{align}
where we neglected the pressure term, while
\begin{align}
\rho_X := \frac{3S q_0 m}{q^2}.
\end{align}
is the energy density scale. We note that $\rho_X$ is inversely proportional to $q^2$. In other words, the equation (\ref{H2effective}) is obtained by taking the Friedmann equation (\ref{FriedmannEq}) and expanding the energy density 
(\ref{EnergyDensity}) (where $S_x$ is a trigonometric function (\ref{Sx}) of $\varphi$ and $\pi_\varphi$) in terms of the standard 
expressions for the scalar field's density and pressure (given by (\ref{rhovarphi}-\ref{Pvarphi})). 
The simplification from (\ref{ModFried}) to (\ref{H2effective}) originates in the late time (oscillatory) evolution of the 
field, which allows us to average out the pressure contribution. Worth mentioning 
is also that the equation (\ref{H2effective}) has a similar structure to the effective Friedmann equation in loop quantum 
cosmology \cite{Bojowald:2008zzb,Ashtekar:2011ni}. 

For the discussed late time dynamics, assuming that the terms ${\cal O}(1/S)$ are still negligible 
and do not affect significantly the approximate solutions, the Friedmann equation (\ref{H2effective}) 
with $\rho_\varphi = c/q$  (where $c$ is some constant) leads to
\begin{align}
H^2 = \frac{\kappa}{3} \frac{c}{q_0 a^3} - \frac{\kappa}{3} \frac{c^2}{3S q_0 m} 
+ {\cal O}(1/S^2)\,, \label{ModFried2}
\end{align}
where we fixed the current value of the scale factor as $a = 1$, which is equivalent to $q = q_0$. 
Surprisingly, while the first contribution to (\ref{ModFried2}) describes the dust matter content 
(which possibly can play a role of dark matter), the second contribution is constant and can be 
interpreted as the negative cosmological constant term, namely
\begin{align}
\Lambda := -\frac{\kappa c^2}{3S q_0 m}\,.
\end{align}
Therefore, within our model both dark matter and negative cosmological constant possibly 
emerge as late time contributions of the scalar field. However, in case the inflationary period
is driven by the field under consideration, its late time contribution to dark matter is naturally 
expected to be marginal. 

Negative cosmological constant may eventually lead to a recollapse, occurring at the density 
scale $\rho_\varphi = \rho_X$. This is under the assumption that the higher order terms will 
not spoil the discussed approximate dynamical behaviour. Writing the density in the form 
$\rho_\varphi = \rho_0 \frac{q_0}{q}$ (such that $c = \rho_0 q_0$), we find that the solution to the 
condition $\rho_\varphi = \rho_X$ is
\begin{align}
q_{collapse} = \frac{3Sm}{\rho_0}\,.
\end{align}
 
Another interesting possibility concerns the slow-roll regime of the field dynamics, in which 
$P_\varphi \approx -\rho_\varphi = {\rm const}$. Then the modified Friedmann equation 
(\ref{ModFried}) simplifies to 
\begin{align}
H^2 = \frac{\kappa}{3} \rho_\varphi - \frac{\kappa}{18} \frac{q^2}{S q_0 m} \rho_\varphi^2 + {\cal O}(1/S^2)\,.
\end{align}
In such a case, the first contribution is approximately constant, while the second one scales as $q^2$. 
The latter corresponds to the effective fluid characterized by the equation of state $P = w_{eff} \rho$, 
with $w_{eff} = -3$, which describes the so-called phantom matter \cite{Caldwell:1999ew}. Consequently, 
the correction term is leading to a recollapse of universe, similarly as in the case of oscillatory regime.  

It is worth stressing that at early times (small $q$) the NFST corrections to the ordinary cosmological model are 
expected to be much smaller than for the late time dynamics. In particular, we 
expect that for sufficiently small $q$ the standard scalar field dynamics is recovered, as one 
can infer from the expansions (\ref{varphieq1/S}) and (\ref{pieq1/S}). The choice of initial 
conditions may, however, be affected by the nontrivial nature of the field phase space. 

On the other hand, there is also a possibility that the negative energy density (\ref{rhonneg}), 
which has been subtracted in the definition of the considered Hamiltonian (\ref{HamSpin2}), 
should actually be taken into account. In such a case it is necessary to balance (\ref{rhonneg}) 
by an additional contribution to the total energy density (e.g.\! radiation or cosmological constant). 
As the result, $\rho_*$ will dominate the energy density at sufficiently small $q$, leading to the 
phase of a cosmic bounce, while at the intermediate energy scales the standard scalar field 
approximation, with $\rho \approx \rho_\varphi$, remains valid. At late times the ${\cal O}(1/S)$ corrections will start to 
prevail, triggering a recollapse. Briefly speaking, the above matter content 
may give us a nonsingular oscillatory cosmological model. This interesting possibility will 
be investigated elsewhere.

\section{Perturbative cosmological inhomogeneities} \label{Pert}

In the previous Section we introduced and discussed the homogeneous cosmological model employing a scalar field 
with the spherical phase space. The tentative analysis of its dynamics, which is determined by a Heisenberg model, 
was performed in the whole range of variability of the field values, including the region far beyond the domain where 
the linear phase space approximation is valid. Our next purpose is to study what are the consequences of applying such 
dynamics at the level of perturbative cosmological inhomogeneities. Similarly as in the homogeneous case, we will 
consider the scalar field theory whose Hamiltonian is derived from the XXZ Heisenberg model. However, while previously 
we calculated the exact form of the evolution equations with an arbitrary $S$ and analyzing them we were using expansions 
up to the first order of $1/S$, here we will restrict to the effects that remain in the $S \rightarrow \infty$ limit. This limit 
corresponds to the quadratic form of the Hamiltonian and the linear order of perturbations. 

Let us again start with an inhomogeneous massive scalar field on the Minkowski spacetime background. Expressing the 
Hamiltonian (\ref{XXZCont}) of the continuous XXZ Heisenberg model in terms of the field variables 
$\varphi$, $\pi_\varphi$, via the relations (\ref{Sx}-\ref{Sz}), and keeping only the leading terms of the 
$1/S$ expansion we obtain
\begin{align}
H_\varphi = \int d^3x \left[ \frac{\pi_\varphi^2}{2} + \frac{1}{2}(\nabla\varphi)^2 + \frac{1}{2} m^2 \varphi^2 + \frac{\Delta}{2m^2} (\nabla\pi_\varphi)^2 \right], 
\label{HamField}
\end{align}
where $m$ is the field's mass, being a function of the parameters of the Heisenberg model (\ref{XXZCont}), 
analogously to (\ref{massed}-\ref{R2def}). The scalar field Hamiltonian (\ref{HamField}) is a generalization 
of the one derived in \cite{Mielczarek:2016xql} from the XXX Heisenberg model, which corresponds to 
$\Delta = 1$. The extra term $\frac{\Delta}{2m^2} (\nabla\pi_\varphi)^2$ for $\Delta=1$ reflects the 
rotational symmetry of the XXX Heisenberg model. Breaking of this symmetry is controlled
by the anisotropy parameter $\Delta$. While for $\Delta \rightarrow 0$ the standard relativistic scalar 
field theory is recovered, for $\Delta = 1$ we obtain the theory (\ref{HamField}) that is invariant under the 
Born reciprocity \cite{Born49} transformation: $\varphi \rightarrow \pi_{\varphi}/m$ and  
$\pi_{\varphi} \rightarrow - \varphi m$.

The detailed discussion of the case $\Delta = 1$ can be found in \cite{Mielczarek:2016xql} and the model 
with an arbitrary value of $\Delta$ will be studied in \cite{XYZSF}. In particular, the spherical phase space 
on which the Hamiltonian (\ref{HamField}) is defined is equipped the Poisson bracket 
\cite{Mielczarek:2016xql} 
\begin{align}
\{f({\bf x}),g({\bf y})\} &= \int \frac{d^3z}{\cos(\pi_\varphi({\bf z}) /R_2)} 
\left( \frac{\delta f({\bf x})}{\delta\varphi({\bf z})} \frac{\delta g({\bf y})}{\delta\pi_\varphi({\bf z})}\right. \nonumber\\ 
&\left.- \frac{\delta f({\bf x})}{\delta\pi_\varphi({\bf z})} \frac{\delta g({\bf y})}{\delta\varphi({\bf z})} \right), 
\label{PoissonBracket}
\end{align} 
which is the field theoretic generalization of the second term of the bracket (\ref{Poisson1}).  

In order to generalize the Hamiltonian (\ref{HamField}) to the FRW background we 
have to apply the following rescalings:
\begin{align}
\pi_\varphi &\rightarrow \frac{\pi_\varphi}{a^3}\,, \\ 
\nabla &\rightarrow \frac{1}{a} \nabla\,, \\ 
d^3x &\rightarrow N a^3 d^3x\,,
\end{align}
where $a$ is the scale factor and $N$ the lapse function. As the result, the scalar field 
Hamiltonian (\ref{HamField}) on the FRW background acquires the form
\begin{align}
H_\varphi &= \int d^3x\, {\cal H}_\varphi  = \int d^3x\, N a^3 \left[ \frac{\pi_\varphi^2}{2a^6} + \frac{1}{2a^2} (\nabla\varphi)^2 \right. \nonumber\\ 
&\left.+ \frac{1}{2} m^2 \varphi^2 + \frac{\Delta}{2m^2 a^8} (\nabla\pi_\varphi)^2 \right]. 
\label{HamFieldFRW}
\end{align}
In what follows we will choose the gauge $N = a$, so that we deal with the conformal time 
$\tau$, which is convenient in the studies of cosmological inhomogeneities. Accordingly, the prime 
``$\ '\ $" will denote a differentiation with respect to the conformal time. 

Our analysis of the cosmological perturbations will also be simplified by 
the assumption that the inhomogeneous part of the field $\varphi$ can be treated as a test field, i.e.\! that it does 
not affect the background dynamics. Furthermore, we do not consider excitations of the gravitational degrees of 
freedom but focus only on the contributions from the matter field. While such an approach is quite 
restrictive, it will give us the first qualitative results concerning the statistical properties of quantum 
cosmological perturbations in a NFST model. The complete analysis, which would take into 
account the scalar gravitational degrees of freedom, is a next step for the future work. 

Instead of using the variable $\varphi$ it is now convenient to introduce its cosmologically rescaled version 
$v := a \varphi$. Then the Hamiltonian density in (\ref{HamFieldFRW}) simplifies to the form 
in which, apart from the term proportional to $\Delta$, gravity manifests only as an effective modification 
of the field's mass (see below). However, due to presence of the non-standard term $\frac{\Delta}{2m^2 a^8}(\nabla\pi_\varphi)^2$, the 
relations between $\varphi$, $\pi_\varphi$ and the momentum $\pi_v$ canonically conjugate 
to $v$ can be expected to differ from the usual cosmological models. Therefore, in order to obtain the Hamiltonian 
$H(v,\pi_v)$ we will apply the following procedure:
\begin{enumerate}
\item We derive the Lagrangian $L(\varphi,\varphi')$ corresponding to 
the Hamiltonian $H(\varphi,\pi_\varphi)$. 
\item We make the change of variables $\varphi = v/a$, which leads to the Lagrangian $L(v,v')$. 
\item Finally, from $L(v,v')$ we calculate the Hamiltonian $H(v,\pi_v)$. 
\end{enumerate}

We begin by finding the equations of motion determined by the Hamiltonian (\ref{HamFieldFRW}):
\begin{align}
\varphi' &= \frac{1}{a^2} \pi_\varphi - \frac{\Delta}{m^2a^4} \nabla^2\pi_\varphi\,, \label{HE1} \\ 
\pi'_\varphi &= a^2 \nabla^2\varphi -m^2 a^4 \varphi\,, \label{HE2}
\end{align}
and we rewrite the first of them as
\begin{align}
\pi_\varphi &= \frac{a^2}{1 - \frac{\Delta}{m^2a^2} \nabla^2}\, \varphi'  \nonumber \\
&= a^2 \left(1 + \frac{\Delta}{m^2 a^2} \nabla^2 \right) \varphi' + {\cal O}(\Delta^2)\,. 
\label{deltapiphi}
\end{align}
The expansion around $\Delta = 0$ introduced here allows us to study deviations from 
the standard relativistic case, in agreement with what was discussed above. We will consider the terms up to the first order 
in $\Delta$. In particular, using (\ref{deltapiphi}) we find that the Hamiltonian density
\begin{align}
{\cal H}_{\varphi} = \frac{\pi_\varphi^2}{2a^2} + \frac{a^2}{2} (\nabla\varphi)^2 
+ \frac{1}{2} m^2 a^4 \varphi^2 + \frac{\Delta}{2m^2 a^4}(\nabla\pi_\varphi)^2 
\label{HamDensPhi}
\end{align}
corresponds to the Lagrangian density
\begin{align}
{\cal L}_{\varphi} &= \frac{a^2}{2} \left((\varphi')^2 - (\nabla\varphi)^2\right) 
- \frac{1}{2} m^2 a^4 \varphi^2  \nonumber \\
&- \frac{\Delta}{2m^2}(\nabla\varphi')^2 + {\cal O}(\Delta^2)\,.   
\end{align}

Proceeding to the second step of our procedure, we change the field variable to $v$, 
which gives us the Lagrangian density
\begin{align}
{\cal L}_v &= \frac{1}{2} \left((v')^2 - (\nabla v)^2\right) - \left(a^2 m^2 - \frac{a^{''}}{a}\right) \frac{v^2}{2} \nonumber \\
&- \frac{\Delta}{2m^2 a^2} \left(\nabla v' - \frac{a'}{a} \nabla v\right)^2 + {\cal O}(\Delta^2) 
\label{Lagv}
\end{align}
and hence we obtain the conjugate momentum
\begin{align}
\pi_v := \frac{\partial{\cal L}_v}{\partial v'} = v' + \frac{\Delta}{m^2 a^2} \nabla^2\left(v' - \frac{a'}{a} v\right) + {\cal O}(\Delta^2)\,. \label{conjugpv}
\end{align}
Therefore, we can calculate that the Hamiltonian density in terms of $v$ and $\pi_v$ has the form
\begin{align}
{\cal H}_v &:= v' \pi_v - {\cal L}_v = \frac{\pi_v^2}{2} + \frac{1}{2} (\nabla v)^2 + \frac{1}{2} m^2_{\text{eff}} v^2 \nonumber \\
&+ \frac{\Delta}{2m^2 a^2} \left(\nabla\pi_v - {\cal H} \nabla v\right)^2 + {\cal O}(\Delta^2)\,,   
\label{Hv}
\end{align}
where ${\cal H} \equiv a'/a$ denotes the conformal Hubble factor and the quantity
\begin{align}
m_{\text{eff}}^2 \equiv m^2 a^2 - \frac{a^{''}}{a}\,.
\end{align}
is the effective mass of the field. 

The equations of motion resulting from (\ref{Hv}) are
\begin{align}
v' &= \pi_v + \frac{\Delta}{m^2 a^2} \left( {\cal H} \nabla^2v - \nabla^2\pi_v \right)+ {\cal O}(\Delta^2)\,, \label{HEv1} \\ 
\pi'_v &= -m_{\text{eff}}^2 v + \left(1 + \frac{\Delta}{m^2 a^2} {\cal H}^2\right) \nabla^2v \nonumber  \\
&- \frac{\Delta}{m^2 a^2} {\cal H} \nabla^2\pi_v + {\cal O}(\Delta^2)\,. \label{HEv2}
\end{align}
Together they lead to the second order equation for $v$:
\begin{align}
& v^{''} - \nabla^2v + m^2_{\text{eff}} v  \nonumber \\
&= \frac{\Delta}{m^2 a^2} \left[ \left(-2{\cal H} + m^2 a^2\right) \nabla^2v + 2{\cal H} \nabla^2v' - \nabla^4v \right], \label{eomv}
\end{align}
where, as usual, the ${\cal O}(\Delta^2)$ terms are neglected. 

However, since it is more convenient to analyze cosmological inhomogeneities in the 
Fourier space, we still have to perform the Fourier transform of the $v$ variable, namely
\begin{align}
v = \int \frac{d^3k}{(2\pi)^{3/2}} v_{\bf k} e^{i {\bf k \cdot x}}\,.
\end{align}
In particular, applying this to (\ref{eomv}) one obtains the following equation of motion for the 
Fourier modes: 
\begin{align}
&v^{''}_{\bf k} + \omega_k^2 v_{\bf k} \nonumber\\ 
&= -\frac{\Delta k^2}{m^2 a^2} \left[2{\cal H}\, v'_{\bf k} + 
\left(k^2 + m^2 a^2 - 2{\cal H}^2\right) v_{\bf k}\right], 
\label{vkEOM}
\end{align}
where $\omega_k^2 \equiv k^2 + m_{\text{eff}}^2$ and $k \equiv \sqrt{\bf k\cdot k}$. 
It will describe evolution of the mode functions in the quantum theory further below.

\subsection{Quantization of modes}

The nonlinear structure of phase space may generally lead to substantial modifications of the field 
quantization procedure (see below). One of the potential problems is that quantum operators should be 
defined for such phase space variables that are globally well defined. In the Appendix we also discuss the 
model in which $R_2$ is a function of $q$ and then the bracket of phase space variables 
is no longer a Poisson bracket, while the corresponding quantum commutator turns out to be non-associative. 
However, since in this Section we remain in the linear limit $S \rightarrow \infty$ of the previous model, 
the standard canonical quantization can still be applied. The only difference with respect to the ordinary 
inflationary theory is then the extra term in the Hamiltonian (\ref{HamFieldFRW}). Therefore, in the current 
Subsection we will perform quantization of the field modes, while in the later ones we will study an impact 
of the nontrivial contributions, controlled by the parameter $\Delta$, on the statistical properties of quantum 
cosmological inhomogeneities. 

We first note that the Poisson bracket between the 
$\varphi$ and $\pi_\varphi$ variables, given by (\ref{PoissonBracket}), can be integrated out to give
\begin{align}
\{\varphi({\bf x}), \pi_\varphi({\bf y})\} = \frac{\delta^{(3)}({\bf x - y})}{\cos(\pi_\varphi({\bf x})/R_2)}\,.
\end{align}
The inverse cosine function on the right hand side strongly affects the quantization, leading in particular 
to a deformation of the commutation relation between field variables \cite{Mielczarek:2016rax}.  
However, since we are restricted here to the linear phase space limit $S \rightarrow \infty$, the Poisson 
bracket of the field variables reduces to the standard canonical expression
\begin{align}
\{\varphi({\bf x}), \pi_\varphi({\bf y})\} = \delta^{(3)}({\bf x - y})\,.
\end{align}
To arrive at the starting point for quantization of our model we switch to the $v$ and $\pi_v$ 
variables, defined in the introductory part of this Section and then the bracket acquires the identical form
\begin{align}
\{v({\bf x}), \pi_v({\bf y})\} = \delta^{(3)}({\bf x - y})\,. \label{PoissPV}
\end{align}
After the canonical quantization $v({\bf x})$ and $\pi_v({\bf y})$ become quantum operators and (\ref{PoissPV}) is replaced by the corresponding commutation relation
\begin{align}
\left[\hat{v}({\bf x}), \hat{\pi}_v({\bf y})\right] = i \delta^{(3)}({\bf x - y})\, \hat{\mathbb{I}}\,.
\label{CommutPV}
\end{align}
Let us remind here that we are using the units in which $\hbar = 1$. 

The operators $\hat{v}({\bf x})$ and $\hat{\pi}_v({\bf y})$ can be now Fourier expanded:
\begin{align}
\hat{v}({\bf x}) &= \int \frac{d^3k}{(2\pi)^{3/2}} \hat{v}_{\bf k} e^{i {\bf k \cdot x}}, \\ 
\hat{\pi}_v({\bf x}) &= \int \frac{d^3k}{(2\pi)^{3/2}} \hat{\pi}_{v \bf k} e^{i {\bf k \cdot x}}.
\end{align}
In the Heisenberg picture we decompose their Fourier modes in the basis of creation and annihilation 
operators  (it can be done because the Hamiltonian operator (\ref{HamQuantV}) introduced below is quadratic):
\begin{align}
\hat{v}_{\bf k}(\tau) &= f_k(\tau)\, \hat{a}_{\bf k} + f_k^*(\tau)\, \hat{a}^\dagger_{-\bf k}\,, \label{vmodes} \\ 
\hat{\pi}_{v\bf k}(\tau) &= g_k(\tau)\, \hat{a}_{\bf k} + g_k^*(\tau)\, \hat{a}^\dagger_{-\bf k}\,, \label{pimodes}
\end{align} 
where $f_k(\tau)$, $g_k(\tau)$ are complex functions, describing the time evolution of 
$\hat{a}^\dagger_{\bf k}$ and $\hat{a}_{\bf k}$. The creation and annihilation operators satisfy the standard 
commutation relation $[\hat{a}_{\bf k}, \hat{a}_{\bf q}^\dagger] = \delta^{(3)}({\bf k} - {\bf q})$ (which would become algebraically deformed in the case of finite $S$ \cite{Mielczarek:2016rax}), while the commutator (\ref{CommutPV}) imposes the requirement that the mode functions are subject to the 
Wronskian condition
\begin{align}
f_k g_k^* - f_k^* g_k = i\,. \label{Wronskianff}
\end{align}

The properly symmetrized quantum version of the Hamiltonian (\ref{HamFieldFRW}), in terms of the Fourier modes $\hat{v}_{\bf k}$, $\hat{\pi}_{v \bf k}$ can be written as
\begin{align}
\hat{H}_v &= \frac{1}{4} \int d^3k \left(1 + \frac{\Delta k^2}{m^2 a^2}\right) 
\left(\hat{\pi}_{v \bf k} \hat{\pi}_{v \bf k}^\dagger + \hat{\pi}_{v \bf k}^\dagger \hat{\pi}_{v \bf k}\right) \nonumber\\ 
&+ \frac{1}{4} \int d^3k \left(\omega^2_k + \frac{\Delta k^2}{m^2 a^2} {\cal H}^2\right) 
\left(\hat{v}_{\bf k} \hat{v}_{\bf k}^\dagger + \hat{v}_{\bf k}^\dagger \hat{v}_{\bf k}\right) \nonumber\\ 
&- \frac{1}{4} \int d^3k\, \frac{\Delta k^2}{m^2 a^2} {\cal H} 
\left( \hat{v}_{\bf k} \hat{\pi}_{v \bf k}^\dagger + \hat{v}_{\bf k}^\dagger \hat{\pi}_{v \bf k} 
+ \hat{\pi}_{v \bf k} \hat{v}_{\bf k}^\dagger + \hat{\pi}_{v \bf k}^\dagger \hat{v}_{\bf k} \right) \nonumber \\
&+ {\cal O}(\Delta^2)\,.
\label{HamQuantV}
\end{align}
Applying the decompositions (\ref{vmodes}-\ref{pimodes}) to 
the Hamilton equation determined by (\ref{HamQuantV}) $\hat{v}_{\bf k}' = -i \left[\hat{v}_{\bf k}, \hat{H}_v\right]$, we derive the equation for mode functions
\begin{align}
f_k' = \left(1 + \frac{\Delta k^2}{m^2 a^2}\right) g_k - \frac{\Delta k^2}{m^2 a^2} {\cal H} f_k + {\cal O}(\Delta^2)\,.
\end{align}
If one solves it for $g_k$, it gives
\begin{align}
g_k = f_k' + \frac{\Delta k^2}{m^2 a^2}({\cal H} f_k - f_k') + {\cal O}(\Delta^2)\,. \label{gsolfk}
\end{align}
Finally, substituting (\ref{gsolfk}) into (\ref{Wronskianff}) we obtain the modified Wronskian 
condition
\begin{align}
f_k (f_k')^* - f_k^* f_k' = i \left(1 + \frac{\Delta k^2}{m^2 a^2} \right) + {\cal O}(\Delta^2)\,, \label{MWronskian}
\end{align}
which has to be satisfied by functions $f_k$.

\subsection{Vacuum normalization}

The choice of the initial state of cosmological inhomogeneities is highly ambiguous. 
In practice, the only way to do it is to assume some particular form of the initial 
state and study its late time behavior. The vacuum state is often perceived as 
a distinguished choice, which is in agreement with the initial homogeneity of the early 
universe. The majority of results for the primordial perturbations have indeed been obtained in the case of 
the initial vacuum state. Therefore, finding the appropriate vacuum state should allow us to 
compare predictions of our model with results of the standard theory of a 
quantum scalar field on the cosmological background. 

The initial vacuum state $|0 \rangle$ is defined as such a state that $\hat{a}_{\bf k} |0 \rangle = 0$ 
(for the time $\tau \rightarrow -\infty$). Furthermore, the vacuum state is assumed 
to be a ground energy state, in which the vacuum expectation value $\langle 0| \hat{H}_v |0\rangle$ 
achieves a minimum. 

In order to find what the ground energy is, we express the quantum Hamiltonian 
(\ref{HamQuantV}) in terms of the creation and annihilation operators, using the decompositions 
(\ref{vmodes}-\ref{pimodes}):
\begin{widetext}
\begin{align}
\hat{H}_v &= \frac{1}{2} \int d^3k \left(\left((1+ A_k) (g_k^*)^2 + (\omega_k^2 + B_k)  (f_k^*)^2 + 2C_k f_k^* g_k^*\right) \hat{a}_{-\bf k}^\dagger 
\hat{a}_{\bf k}^\dagger \right.\nonumber\\
&+ \left((1+ A_k) g_k^2 + (\omega_k^2 + B_k) f_k^2 + 2C_k f_k g_k\right) \hat{a}_{\bf k} \hat{a}_{-\bf k} \nonumber\\
&+ \left((1+ A_k) |g_k|^2 + (\omega_k^2 + B_k) |f_k|^2 + 2C_k f_k^* g_k\right) \hat{a}_{-\bf k}^\dagger \hat{a}_{-\bf k} \nonumber\\
&\left.+ \left((1+ A_k) |g_k|^2 + (\omega_k^2 + B_k) |f_k|^2 + 2C_k f_k g_k^*\right) \hat{a}_{\bf k} \hat{a}_{\bf k}^\dagger\right), 
\label{Haadag}
\end{align}
\end{widetext}
where we denoted $A_k \equiv \frac{\Delta k^2}{m^2 a^2}$, $B_k \equiv 
\frac{\Delta k^2}{m^2 a^2} {\cal H}^2$ and $C_k \equiv \frac{\Delta k^2}{m^2 a^2} {\cal H}$. 
Furthermore, as justified earlier the standard commutation relation $[\hat{a}_{\bf k}, \hat{a}_{\bf q}^\dagger] = 
\delta^{(3)}({\bf k} - {\bf q})$ can be used. Consequently, the vacuum expectation value of (\ref{Haadag}) is calculated to be
\begin{align}
\langle 0| \hat{H}_v |0\rangle &= \frac{1}{2} \delta^{(3)}(0) \int d^3k \left((1 + A_k) |g_k|^2\right. \nonumber\\ 
&\left.+ (\omega_k^2 + B_k) |f_k|^2 + 2C_k f_k g_k^*\right)
\end{align}
and we identify
\begin{align}
E_k \equiv (1 + A_k) |g_k|^2 + (\omega_k^2 + B_k) |f_k|^2 + 2C_k f_k g_k^* \label{Ek}
\end{align}
as the energy density of a given mode $k$. 

Let us now introduce the polar decomposition of each complex function $f_k = r_k e^{i \alpha_k}$. 
From the Wronskian condition (\ref{MWronskian}) we obtain the relation $\alpha_k' = -A_k/(2r_k^2)$, which can be used 
to eliminate $\alpha_k$ from (\ref{Ek}), so that it becomes
\begin{align}
E_k &= (1 - A_k) r'^2 + 4C_k r' r + (\omega_k^2 + B_k) r^2 \nonumber \\
&+ (1 + A_k) \frac{1}{4r^2} + i C_k + {\cal O}(\Delta^2)\,.
\end{align}
Subsequently, to find a minimum of $E_k$ we calculate its derivatives
\begin{align}
\frac{1}{\partial r'} E_k &= 2 (1 - A_k) r' + 4C_k r\,, \\
\frac{1}{\partial r} E_k &= 4C_k r' + 2 (\omega_k^2 + B_k) r - (1 + A_k) \frac{1}{2r^3}
\end{align}
and set both these expressions to zero. The solution of the resulting system of equations is given by
\begin{align}
r_k' &= -2{\cal H} \frac{\Delta k^2}{m^2 a^2} r_k + {\cal O}(\Delta^2)\,, \\
r_k &= \frac{1}{\sqrt{2\omega_k}} \left[ 1 + \frac{1}{4} \frac{\Delta k^2}{m^2 a^2} 
\left(1 - \frac{{\cal H}^2}{\omega_k^2}\right) + {\cal O}(\Delta^2) \right] \label{rsol}
\end{align}
(it should be stressed that $\omega_k$ depends on time via $m^2_{\text{eff}}$) and consequently we also obtain
\begin{align}
\alpha_k &= -\int d\tau\, \omega_k \left[1 + \frac{1}{2} \frac{\Delta k^2}{m^2 a^2} 
\left(1 + \frac{{\cal H}^2}{\omega_k^2}\right)\right] \nonumber\\
&+ {\cal O}(\Delta^2)\,. \label{asol}
\end{align}
In the UV limit $k^2 \gg m_{\text{eff}}^2$, and also assuming the condition $\frac{\Delta k^2}{m^2 a^2} \ll 1$, 
we can use (\ref{rsol}) and (\ref{asol}) to write the mode function in the form
\begin{align}
f_k &= \frac{1}{\sqrt{2k}}\left[1 + \frac{1}{4} \frac{\Delta k^2}{m^2 a^2}\right] 
\exp\left[-i k \tau - i \frac{\Delta k^3}{m^2} \int \frac{d\tau}{a^2}\right] \nonumber\\
&+ {\cal O}(\Delta^2) \nonumber\\
&= \frac{e^{-i k \tau}}{\sqrt{2k}}\left[1 + \frac{\Delta k^2}{m^2 a^2} \left(\frac{1}{4} - i k a^2 \int \frac{d\tau}{a^2}\right)\right] \nonumber\\
&+ {\cal O}(\Delta^2)\,, \label{DeltaBD}
\end{align}
which describes the $\Delta$-modified version of the Bunch-Davies vacuum state.
 
\subsection{Inflationary power spectrum}

The Bunch-Davies vacuum normalization of the mode functions derived in the previous 
Subsection allows us to quantify the statistical properties of the vacuum field configuration. 
For the linear inhomogeneities, as the ones considered in this Section, the vacuum expectation 
values of the products of the physical field operators $\hat{\varphi} := \hat{v}/a$ 
carry the whole necessary information about correlations of quantum states. In particular, 
the two-point correlation function can be written in the form
\begin{align}
\langle 0| \hat{\varphi}({\bf x},\tau) \hat{\varphi}({\bf y},\tau) |0 \rangle 
= \int_0^\infty \frac{dk}{k} \frac{\sin kr}{kr} {\cal P}_\varphi(k,\eta)\,, 
\label{CorrPower} 
\end{align}
where the power spectrum is defined as
\begin{align}
{\cal P}_\varphi(k,\tau) := \frac{k^3}{2\pi^2} \left|\frac{f_k(\tau)}{a(\tau)}\right|^2 
\label{Pspectrum}
\end{align}
and we denote $r = |{\bf x} - {\bf y}|$. 

As one can verify, the evolution equation for the mode functions $f_k$ has the same form as the 
equation for the Fourier modes (\ref{vkEOM}), namely
\begin{align}
&f^{''}_k + 2 {\cal H} \frac{\Delta k^2}{m^2 a^2} f'_k \nonumber \\
&+ \left[\omega_k^2 + \frac{\Delta k^2}{m^2 a^2} \left(k^2 + m^2 a^2 - 2{\cal H}^2\right) \right] f_k = 0\,. 
\label{modeeq}
\end{align}
 We notice that in the expanding regime (when the conformal Hubble factor ${\cal H} < 0$) the 
modes are additionally dumped by the negative ``friction term", proportional to ${\cal H}$. This term 
can be eliminated from the equation by introducing the new variable
\begin{align}
y_k := \exp\left(-\frac{1}{2} \frac{\Delta k^2}{m^2 a^2}\right) f_k\,,
\end{align}
in which (\ref{modeeq}) becomes
\begin{align}
y^{''}_k + \left[ \omega_k^2 \left(1 + \frac{\Delta k^2}{m^2 a^2}\right) + \frac{\Delta k^2}{m^2 a^2} {\cal H}^2 \right] y_k = 0\,. 
\label{eqyk}
\end{align}

As a specific example let us consider the de Sitter background, which is the leading order 
approximation (vanishing slow-roll parameters) of the inflationary period in cosmology. 
Such a period is expected to occur due to the background dynamics studied in Sec.~\ref{Hom}. 
It is known that in the de Sitter phase the time dependence of the scale factor is given by
\begin{align}
a = -\frac{1}{H \tau}\,, 
\label{aDeSitter}
\end{align}
with the constant Hubble factor $H$. Consequently, making the change of variables to 
$x = -k \tau$ we can express the equation for $y_k$ 
as
\begin{align}
y^{''}_k + \Omega^2(x)\, y_k = 0\,, 
\label{Eqyk}
\end{align}
with the following time-dependent frequency squared factor:  
\begin{align}
\Omega^2(x) := \left(1 - \frac{2 - 3\eta}{x^2}\right) \left(1 + \frac{\Delta}{3\eta} x^2\right) + \frac{\Delta}{3\eta}\,,
\end{align}
where we introduced the dimensionless parameter $\eta \equiv \frac{m^2}{3H^2}$. The equation 
(\ref{Eqyk}) does not have a simple solution. However, there is actually no need to solve it in the full 
domain of $x$. From the perspective of cosmological observations only the super-Hubble regime, 
where $x \ll 1$, is relevant. We observe that in this regime there is no contribution proportional to $\Delta$ to the scaling 
of $\Omega^2(x)$ and therefore $\Omega^2(x) \approx -\frac{2 - 3\eta}{x^2}$, which is the same as 
in the ordinary theory. Consequently, the growing mode solution to (\ref{Eqyk}) is $y_k \propto x^{-1 + \eta/2}$, 
which corresponds to $f_k \propto \exp\left(\frac{1}{2} \frac{\Delta x^2}{3\eta}\right) x^{-1 + \eta/2}$. The only difference 
with respect to the standard case is the exponential factor, which, however, does not affect the power-low 
dependence and in consequence the spectral index. The exponential factor leads to a slight 
shift of the amplitude of perturbations and agrees with the contribution expected from the $\Delta$-modified 
version of the Bunch-Davies vacuum derived in the previous Subsection. 

Namely, for the de Sitter background dynamics (\ref{aDeSitter}) the $\Delta$-modified Bunch-Davies vacuum 
defined in (\ref{DeltaBD}) reduces to
\begin{align}
f_k &= \frac{e^{i x}}{\sqrt{2k}} \left[ 1 + \frac{\Delta x^2}{3\eta} \left(\frac{1}{4} + \frac{i}{x^2} \int x^2 dx\right) \right] \nonumber\\
&+ {\cal O}(\Delta^2)\,,
\end{align}
for which the power spectrum 
\begin{align}
{\cal P}_\varphi(k,\tau) = \left(\frac{H}{2\pi}\right)^2 x^2 \left(1 + \frac{\Delta}{6\eta} x^2 + {\cal O}(\Delta^2) \right)
\end{align}
and at the horizon scale $x \approx 1$ it gives 
\begin{align}
{\cal P}_\varphi(x \approx 1) \approx \left(\frac{H}{2\pi}\right)^2 \left( 1 + \frac{\Delta}{6\eta} + {\cal O}(\Delta^2) \right). 
\end{align}
The magnitude of derived corrections is inversely proportional to $\eta$, and therefore $\eta$ has to be 
sufficiently large in order to avoid deviations from the known results. More precisely, the ratio $m/H$ has to satisfy the following consistency condition $m/H \gg \sqrt{\Delta}$. 

On the basis of the above analysis one can conclude that the spectral index
\begin{align}
n_S := \frac{d\ln {\cal P}_\varphi(x = 1)}{d\ln k} = 0
\end{align} 
has no leading order deviations in $\Delta$. In other words, the power spectrum remains scale-invariant, 
as expected for the de Sitter phase. The presence of $\Delta$ is manifest only in the amplitude of perturbations. 
However, the higher order corrections in $\Delta$ will have a non-vanishing contribution, similarly as the 
corrections linear in $\Delta$ but multiplied by the slow-roll parameters, which is not considered in the lowest 
order discussion presented here.

\section{Summary} \label{Sum}

This paper provides the first attempt to apply the recently introduced Nonlinear Field 
Space Theory (NFST) to the domain of cosmology. Let us stress that general relativity itself 
was not modified but we focused our attention on a scalar field describing the matter content 
of the standard cosmological model. The field was generalized to have the spherical phase space, 
on which we defined the appropriate symplectic form. In principle, other choices for a bilinear 
two-form on this phase space are possible as well. Some results for one of such forms are 
discussed in the Appendix below. 

Using the analogy between a scalar NFST with the spherical phase space and a system of spins, 
we borrowed the Hamiltonian for the matter field from the XXZ Heisenberg model and adapted it to 
the FRW background. Then our considerations were restricted to a homogeneous cosmological model. 
As it was shown, the standard dynamics of a massive scalar field is recovered for sufficiently 
small volumes of universe. On the other hand, it was found that at late times the effects of NFST may 
become significant. Observational implications of this possibility and a detailed analysis of the discussed 
model deserve to be the subject of further investigations. The preliminary results suggest that the phase 
of a cosmic bounce, which replaces the big bang singularity, can also be obtained within our framework. 

Subsequently, we studied a generation of primordial quantum inhomogeneities in the scalar field theory corresponding 
to the XXZ Heisenberg model. The leading order contributions of the anisotropy parameter 
$\Delta$ were investigated. However, for the considered linear order of the perturbative analysis, the effects 
of nonlinearity of the field phase space were not taken into account. Such higher-order effects are 
unavoidably associated with the non-Gaussian features. Since non-linearity is the inherent feature of the 
NFST proposal, studies of the non-Gaussianity within this framework may provide a powerful tool 
to confront the predictions of NFST with the cosmological data. 

It was shown that no corrections to the spectral index are expected at the linear order in $\Delta$. 
However, it has to be stressed that we adopted certain simplifications in our calculations. In particular, 
a decomposition of the NFST scalar field into the background and perturbation contributions has 
to be investigated. Due to the field dependent function in the Poisson bracket, such a decomposition of 
the kinematics will require a subtle treatment. Only after it is done, the homogenous background field 
can be considered a source of dynamics, on the top of which the inhomogeneous modes are introduced. 
In our simplified analysis we have not extracted the zero mode from the field dynamics, and assumed that 
$\langle 0| \hat{\varphi}({\bf x},\tau) |0 \rangle = 0$. 

Worth stressing is that the results of this paper open a novel possibility of building relations 
between cosmology and condensed matter physics, thanks to the duality between spin systems and 
NFST with the spherical phase space that has been discussed in the Introduction. In particular, the 
$\Delta \rightarrow 0$ case, which is the relativistic scalar NFST, is dual to the so-called XY model, 
which normally provides a description of the superconductive state of matter \cite{KT73}. This relationship 
may turn out to be a source of new ideas for both cosmology and condensed matter physics. 
It concerns not only models with a scalar field, but also with other types of fields (such 
as spinor and gauge fields), for which the condensed matter dual descriptions can be potentially 
introduced. 

\section*{Acknowledgements}

J.M. is supported by the National Science Centre Poland project DEC-2013/09/B/ST2/03455. 
J.M. and T.T. have additionally been supported by the Iuventus Plus grant 0302/IP3/2015/73 
from the Polish Ministry of Science and Higher Education, and T.T. by the National Science 
Centre Poland, project 2014/13/B/ST2/04043.

\section*{Appendix}

Let us here consider the model in which we do not assume that the two-form (\ref{SympNew}) on the 
total phase space has to be a closed form. Then to identify the standard Hamiltonian (\ref{HamStandard2}) 
as the small field limit of (\ref{HamSpin1}), under the assumption that $R_1 R_2 \rightarrow S$ for 
$q \rightarrow q_0$, we need to impose the relations:
\begin{align}
\tilde{\mu} B_x &= \frac{m}{q}\,, \label{massedA}\\ 
R_1 &= \sqrt{\frac{S}{qm}}\,, \label{R1defA}\\ 
R_2 &= \sqrt{Sqm}\,, \label{R2defA}
\end{align}
which give us simply $R_1 R_2 = S = {\rm const}$. In this case $q_0$ does not appear in any formulae. 
Similarly as in (\ref{HamSpin2}), taking into account the appropriate (constant) energy shift we obtain the proper Hamiltonian
\begin{align}
H_{\bf S} &= N m \left(S - S_x\right) \nonumber \\
&= N q \left(\frac{\pi^2_\varphi}{2q^2} + \frac{1}{2} m^2 \varphi^2\right) + {\cal O}(4)\,. 
\label{HamSpin2A}
\end{align}
Consequently, the expression for energy density (\ref{EnergyDensity}) becomes
\begin{align}
\rho = \frac{m}{q} \left(S - S_x\right) 
\label{EnergyDensityA}
\end{align}
and the expansion of the (modified) Friedmann equation in $1/S$ can be written as
\begin{align}
H^2 = \frac{\kappa}{3} \rho_\varphi - \frac{\kappa}{9} \frac{q}{Sm}
\left(\rho_\varphi^2 - \frac{1}{2} P_\varphi^2 \right) + {\cal O}(1/S^2)\,. \label{ModFriedA}
\end{align}
which differs from (\ref{ModFried}) only by the absent factor $q/q_0$ in the first order term. 

The two-form (\ref{SympNew}) with $R_2(q)$ given by (\ref{R2defA}) is
\begin{align}
\omega_{\varphi,q} &:= \cos\left(\frac{\pi_\varphi}{R_2(q)}\right) d\pi_\varphi \wedge d\varphi \nonumber \\
&= d\pi_\varphi \wedge d\varphi + {\cal O}\left(R_2^{-4}(q)\right). \label{SympNewA}
\end{align}
As one can observe, in the $R_2 \rightarrow \infty$ limit $\omega_{\varphi,q}$ reduces to the standard 
symplectic form $\omega = d\pi_\varphi \wedge d\varphi$. However, (\ref{SympNewA}) is not a closed 
form, $d\omega_{\varphi,q} \neq 0$. This has a major impact on the corresponding algebra of phase 
space variables, whose bracket is determined by inverse of the total form $\omega_{\text{FRW}} + \omega_{\varphi,q}$ and given by
\begin{align}
\left\{\cdot,\cdot\right\}_{\varphi,q} &= \left[ \frac{\partial\cdot}{\partial q} \frac{\partial\cdot}{\partial p} 
- \frac{\partial\cdot}{\partial p} \frac{\partial\cdot}{\partial q} \right] \nonumber \\
&+ \frac{1}{\cos\left(\frac{\pi_\varphi}{R_2(q)}\right)}\left[ \frac{\partial\cdot}{\partial\varphi} 
\frac{\partial\cdot}{\partial\pi_\varphi} 
- \frac{\partial\cdot}{\partial\pi_\varphi} \frac{\partial\cdot}{\partial\varphi} \right].
\label{Poisson1A}
\end{align}
The above algebra does not satisfy the Jacobi identity and therefore is not a Poisson algebra, as well as 
can not become the associative algebra of operators after quantization. Namely, for arbitrary functions 
on phase space $f$, $g$ and $h$, the so-called Jacobiator of the bracket (\ref{Poisson1A}) is non-zero and has the form
\begin{widetext}
\begin{align}
\{f,g,h\} &:= \{f,\{g,h\}\} + \{h,\{f,g\}\} + \{g,\{h,f\}\} \nonumber\\ 
&= \frac{\pi_\varphi}{2q \sqrt{Smq}} \frac{\sin\frac{\pi_\varphi}{\sqrt{Smq}}}{\cos^2\frac{\pi_\varphi}{\sqrt{Smq}}} \left(\frac{\partial f}{\partial p}\, [g,h]_\varphi + \frac{\partial h}{\partial p}\, [f,g]_\varphi + \frac{\partial g}{\partial p}\, [h,f]_\varphi\right) \nonumber \\ 
&= \left(\frac{\pi_\varphi^2}{2Sm q^2} + {\cal O}(1/S^2)\right) \left(\frac{\partial f}{\partial p}\, [g,h]_\varphi + \frac{\partial h}{\partial p}\, [f,g]_\varphi + \frac{\partial g}{\partial p}\, [h,f]_\varphi\right),
\end{align}
\end{widetext}
where $[\cdot,\cdot]_\varphi \equiv \frac{\partial\cdot}{\partial\varphi} 
\frac{\partial\cdot}{\partial\pi_\varphi} 
- \frac{\partial\cdot}{\partial\pi_\varphi} \frac{\partial\cdot}{\partial\varphi}$. This is not a problem by itself, since such 
non-Poisson physical systems have already been considered from the theoretical perspective and can actually exist 
in nature, see e.g.\! \cite{Bojowald:2014oea} and references therein. Moreover, they may lead to appearance of the 
fundamental length \cite{Stueckelberg:1960qe} (see \cite{Gunaydin:2013ny} for a relation with string theory). These 
systems require a special treatment, especially regarding quantization, where one has to use the *-products or other 
refined constructions. However, in our investigations of the quantum regime in Sec.~\ref{Pert} we restricted to the limit 
$S \rightarrow \infty$, where the form (\ref{SympNewA}) becomes a (closed) symplectic form. The remaining non-standard 
features of the theory are then a consequence of non-zero parameter $\Delta$ in the scalar field Hamiltonian and therefore 
the results of Sec.~\ref{Pert} are also valid in the current case. 

We note that a violation of the Jacobi identity in (\ref{Poisson1A}) is associated only with the functions that depend on $p$. 
Furthermore, the brackets between any pair of phase space variables $S_x,S_y,S_z$, $q$ and $p$ do not change except
\begin{align}
\{S_x,p\} & = \frac{1}{2q} \left(-S_y \arctan\frac{S_y}{S_x} + \frac{S_x S_z}{\sqrt{S^2 - S_z^2}}\, \arcsin\frac{S_z}{S}\right), \\   
\{S_y,p\} & = \frac{1}{2q} \left(S_x \arctan\frac{S_y}{S_x} + \frac{S_y S_z}{\sqrt{S^2 - S_z^2}}\, \arcsin\frac{S_z}{S}\right), \\  
\{S_z,p\} & = -\frac{1}{2q} \sqrt{S^2 - S_z^2}\, \arcsin\frac{S_z}{S}\,.
\end{align}
Consequently, now the evolution equations for the $S_x,S_y,S_z$ variables become
\begin{widetext}
\begin{align}
\dot{S}_x &= \{S_x, H_{\text{tot}}\} = -\frac{3}{4} N \kappa p \left(-S_y \arctan\frac{S_y}{S_x} + \frac{S_x S_z}{\sqrt{S^2 - S_z^2}}\, \arcsin\frac{S_z}{S}\right), \label{eqSxA} \\ 
\dot{S}_y &= \{S_y, H_{\text{tot}}\} = N m\, S_z - \frac{3}{4} N \kappa p \left(S_x \arctan\frac{S_y}{S_x} + \frac{S_y S_z}{\sqrt{S^2 - S_z^2}}\, \arcsin\frac{S_z}{S}\right), \label{eqSyA} \\ 
\dot{S}_z &= \{S_z, H_{\text{tot}}\} = -N m\, S_y + \frac{3}{4} N \kappa p \sqrt{S^2 - S_z^2}\, \arcsin\frac{S_z}{S} \label{eqSzA}
\end{align}
\end{widetext}
and the equation for $p$ is
\begin{align}
\dot{p} &= \frac{3}{4} N \kappa p^2 + N m \frac{1}{2q} \left(-S_y \arctan\frac{S_y}{S_x}\right. \nonumber\\ 
&\left.+ \frac{S_x S_z}{\sqrt{S^2 - S_z^2}}\, \arcsin\frac{S_z}{S}\right). \label{eqpA}
\end{align}
On the other hand, the equations (\ref{eqphi}-\ref{eqpi}) in the current case have the identical form as before but with the implicit expressions for $R_1$ and $R_2$ given by (\ref{R1defA}-\ref{R2defA}). The latter feature manifests itself if we expand the equations up to the first order in $1/S$, obtaining
\begin{align}
\dot{\varphi} &= \frac{\pi_\varphi}{q} + \frac{\pi_\varphi}{Sm} 
\left(\frac{\pi_\varphi^2}{3q^2} - \frac{m^2}{2} \varphi^2\right) + {\cal O}(1/S^2)\,, \\ 
\dot{\pi}_\varphi &= -q m^2\, \varphi + \frac{q^2 m^3}{6S} \varphi^3 + {\cal O}(1/S^2)\,,
\end{align}
which differ with respect to (\ref{varphieq1/S}-\ref{pieq1/S}) by the absent factor $q/q_0$ in the first order terms. From the above equations we also derive the corresponding modified Klein-Gordon equation
\begin{align}
& \ddot{\varphi} + 3H \dot{\varphi} + m^2 \varphi = \nonumber\\
&-\frac{qm}{S} \left[ \frac{H}{m^2} \dot{\varphi}^3 + \frac{3H}{2} \dot{\varphi} \varphi^2 + 2\, \dot{\varphi}^2 \varphi - \frac{2m^2}{3} \varphi^3 \right] \nonumber\\
&+ {\cal O}(1/S^2)\,,
\end{align}
which has one additional term in comparison with (\ref{KGeq}).

\end{document}